\documentclass[twocolumn]{aastex631}
\usepackage{amsmath}
\usepackage{graphicx}
\usepackage{comment}
\newcommand{\alf}{Alfv\'{e}n\,}
\newcommand{\Rs}{$R_{\odot}$\,}
\newcommand{\PF}{$(S_{A}/B)_{\odot}$\,}

\newcommand{\nss}[1]{\textcolor{purple}{\textit{#1}}}

\submitjournal{ApJ}
\shorttitle{Solar wind modeling}
\shortauthors{Sachdeva et al.}
\submitjournal{ApJ}

\begin{document}
\title{Solar wind modeling with the \alf Wave Solar atmosphere Model driven by HMI-based Near-Real-Time maps by the National Solar Observatory}
\author[0000-0001-9114-6133]{Nishtha Sachdeva} 
\affiliation{Department of Climate and Space Sciences and Engineering, University of Michigan, Ann Arbor, MI 48109, USA}
\author[0000-0003-0472-9408]{Ward B. Manchester IV} 
\affiliation{Department of Climate and Space Sciences and Engineering, University of Michigan, Ann Arbor, MI 48109, USA}
\author[0000-0002-6118-0469]{Igor Sokolov} 
\affiliation{Department of Climate and Space Sciences and Engineering, University of Michigan, Ann Arbor, MI 48109, USA}
\author[0000-0003-1674-0647]{Zhenguang Huang}
\affiliation{Department of Climate and Space Sciences and Engineering, University of Michigan, Ann Arbor, MI 48109, USA}
\author[0000-0001-9746-9937]{Alexander Pevtsov}
\affiliation{National Solar Observatory, 3665 Discovery Drive, 3rd Floor Boulder, CO 80303 USA}
\author[0000-0002-1155-7141]{Luca Bertello}
\affiliation{National Solar Observatory, 3665 Discovery Drive, 3rd Floor Boulder, CO 80303 USA}
\author[0000-0003-0489-0920]{Alexei A. Pevtsov}
\affiliation{National Solar Observatory, 3665 Discovery Drive, 3rd Floor Boulder, CO 80303 USA}
\author[0000-0001-8459-2100]{Gabor Toth}
\affiliation{Department of Climate and Space Sciences and Engineering, University of Michigan, Ann Arbor, MI 48109, USA}
\author[0000-0001-5260-3944]{Bart van der Holst}
\affiliation{Department of Climate and Space Sciences and Engineering, University of Michigan, Ann Arbor, MI 48109, USA}
\correspondingauthor{Nishtha Sachdeva}
\email{nishthas@umich.edu}

\begin{abstract}
We explore model performance for the \alf Wave Solar atmosphere Model (AWSoM) with near-real-time (NRT) synoptic maps of the photospheric vector magnetic field. These maps, produced by assimilating data from the Helioseismic Magnetic Imager (HMI) onboard the Solar Dynamics Observatory (SDO), use a different method developed at the National Solar Observatory (NSO) to provide a near contemporaneous source of data to drive numerical models. Here, we apply these NSO-HMI-NRT maps to simulate three Carrington rotations (CRs): 2107
-2108 (centered on 2011/03/07 20:12 CME event), 2123 (integer CR) and 2218--2219 (centered on 2019/07/2 solar eclipse), which together cover a wide range of activity level for solar cycle 24. We show simulation results, which reproduce both extreme ultraviolet emission (EUV) from the low corona while simultaneously matching {\it in situ} observations at 1 au as well as quantify the total unsigned open magnetic flux from these maps.
\end{abstract}

\section{Introduction} \label{sec:intro}
Global estimates of the solar photospheric magnetic field in the form of synoptic and synchronic maps are the fundamental empirical data product that allow for simulation and prediction of the three-dimensional (3D) structure of the solar corona, solar wind, and the heliosphere \citep{Mikic:1999, Roussev:2003a, Usmanov:2003, Cohen:2007, vanderHolst:2010, Lionello:2013, Sokolov:2013, vanderHolst:2014, Riley:2014, Feng:2014, Feng:2015, Riley:2019, vanderHolst:2019}. These maps of the photospheric magnetic field are constructed from a time-series of full-disk magnetograms (collected over a solar rotation period of 27 days or more), which are then modified and assembled to simultaneously cover the entire solar surface. Photospheric full-surface maps became available shortly following the routine production of full-disk magnetograms, beginning with the Global Oscillation Network Group (GONG) \cite[see, e.g.][]{Donaldson:2002}. The Stanford approach for producing synoptic maps from the Helioseismic Magnetic Imager (HMI) data is to use only 2 degree strips of data at the solar central meridian from each full disk magnetogram and stitch these strips together to form synoptic maps.

Perhaps the most advanced system for producing global photospheric maps is the Air Force Data Assimilative Photospheric flux Transport (ADAPT) model, which is a flux transport model that makes use of data assimilation for incorporating magnetic field data. In ADAPT, the photospheric magnetic flux is transported by differential rotation, meridional flows and convection-driven diffusion while observational data-driven updates to the model are made using data assimilation techniques \citep{Arge:2000, Arge:2013}. ADAPT maps are routinely used in numerical simulations, including our model validation work \citep{Sachdeva:2019, Sachdeva:2021} and simulations of Parker Solar Probe encounters \citep{vanderHolst:2019, vanderHolst:2022psp} using AWSoM. Currently NSO provides ADAPT maps driven with GONG magnetograms (\url{https://gong.nso.edu/adapt/maps/}).

While the use of HMI and GONG informed ADAPT maps has been extremely successful, these data products are not suitable for use in near-real-time (NRT) simulations as a result of the significant time delay in producing the maps. For space weather forecasting, accurate maps with minimum delay from the moment the magnetic fields are observed are required. For this purpose, the National Solar Observatory embarked on a mission to produce NRT synoptic maps specifically designed as input for numerical models to forecast the coronal space environment. The synoptic map data products are available via \textcolor{blue}{doi: 10.25668/nw0t-b078}.  

The NSO approach is to speed map creation by using the full disk vector magnetogram, and weight pixel contribution based on its distance from the central meridian (see, \citet{Bertello:2014}). The maps (hereafter referred to as NSO-HMI-NRT) are a product of this NSO approach applied to HMI full-disk magnetograms. SOLIS/VSM vector data may also be used for NRT maps. While HMI and SOLIS/VSM produce different results for weaker fields and sometimes show the opposite orientation in transverse fields \citep{Pevtsov.etal2021,Liu.etal2022}, the two instruments agree very well in strong field regions \citep{Pietarila.etal2013,Riley:2014}. The disagreement in weak magnetic field regions is not the result of disambiguation, but mostly due to differences in noise levels and magnetic fill factor (fraction of magnetized and non-magnetized plasma contribution to a single pixel) \citep{Pevtsov.etal2021}. 

In this work, we explore the performance of AWSoM driven with NSO-HMI-NRT maps. For this goal, we choose three Carrington rotations (CRs): 2107--2108 (centered on 2011/03/07 20:12 UT, a CME event), 2123 (integer CR) and 2218--2219 (centered on 2019/07/2, total solar eclipse), which cover the ascending phase, solar maximum and solar minimum of the solar cycle. For simplicity, hereafter we refer to these synoptic maps using their nearest integer rotation number (i.e., CR2107, CR2123, and CR2219) although two of them straddle more than one Carrington rotation. We then make direct comparisons to observed data to provide a measure of model fidelity, first for coronal images made in the extreme ultra violet, and second with {\it in situ} time series data extracted near Earth. This two-type data comparison is less thorough than previous model validation efforts \citep{Cohen:2007, Jin:2012, Sachdeva:2019, Sachdeva:2021}, but will serve the purpose of demonstrating model performance with the maps designed for space weather forecasting. In Sections \ref{sec:awsom} and \ref{sec:maps}, we briefly describe the AWSoM model and the NSO-HMI-NRT maps while Section \ref{sec:setup} describes the simulation design. Sections \ref{sec:results} and \ref{sec:conclude} describe simulation results and summarize this work.

\section{ALFV\'{E}N Wave Solar atmosphere Model (AWSoM)}\label{sec:awsom}
AWSoM \citep{vanderHolst:2014awsom,Sokolov:2013} within the Space Weather Modeling Framework (SWMF; \citet{Toth:2012swmf}) is a self-consistent, 3D global magnetohydrodynamic (MHD) model with its inner boundary at the base of the transition region (upper chromosphere) extending into the solar corona and the heliosphere. It is driven by the radial component of the photospheric magnetic field at the inner boundary. Like most solar corona models, this input comes from the solar synoptic/synchronic magnetic field maps, which is essential for reliable predictions. AWSoM incorporates the low-frequency \alf wave turbulence as a consequence of the non-linear interaction of forward and counter propagating \alf waves, which is based on well established theories describing the evolution and transport of \alf turbulence, (e.g., \citet{Hollweg:1986, Matthaeus:1999, Zank:2014, Zank:2017}). The AWSoM phenomenological approach self-consistently describes the heating and acceleration of the solar wind in response to turbulence while not yet including many higher-order physical effects. Several other extended MHD coronal models have been developed \citep{Usmanov:2000, Suzuki:2005, Lionello:2014}, which also include \alf wave turbulence. AWSoM is distinguished from other global MHD models by including proton temperature anisotropy (perpendicular and parallel ion temperature), isotropic electron temperature, heat conduction and radiative cooling. The wave dissipation heats the solar wind plasma and the (thermal and nonthermal) pressure gradients accelerate the solar wind \citep{Meng:2015}. The full set of MHD equations using the Block Adaptive Tree Solarwind-Roe-Upwind Scheme (BATS-R-US; \citet{Powell:1999}) numerical scheme are solved within AWSoM. A detailed description of the model equations and their implementation is available in \citet{vanderHolst:2014awsom}. The energy partitioning scheme in AWSoM has been significantly improved and recently validated against the data from Parker Solar Probe \citep{vanderHolst:2019,vanderHolst:2022psp}. These improvements include using the critical balance formulation of \citet{Lithwick:2007} and implementation of the alignment angle between the counter-propagating \alf waves in the energy cascade.

\begin{figure*}
\gridline{ \fig{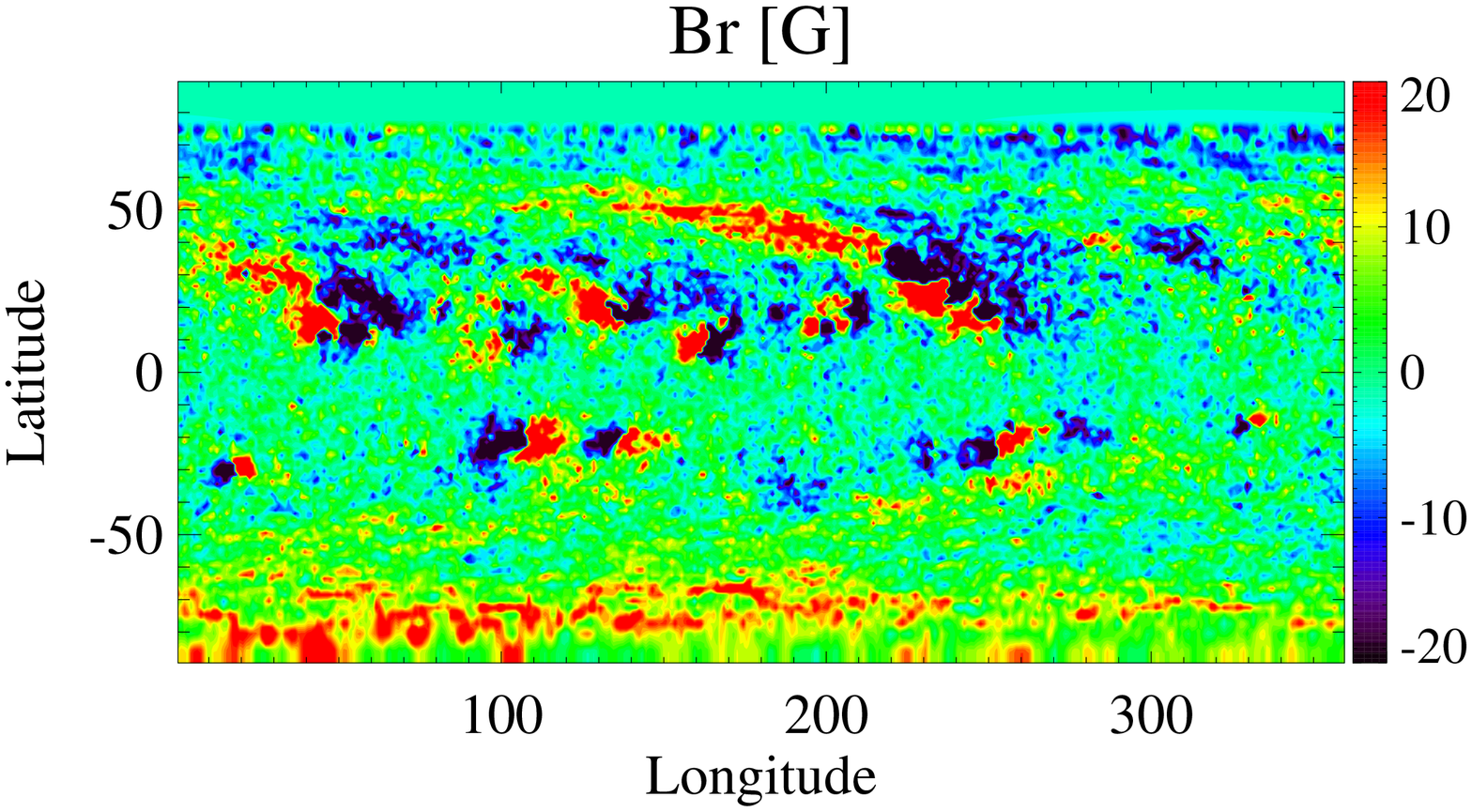}{0.5\textwidth}{\vspace{-1.0cm}(a) CR2107 NSO-HMI-NRT map}
\fig{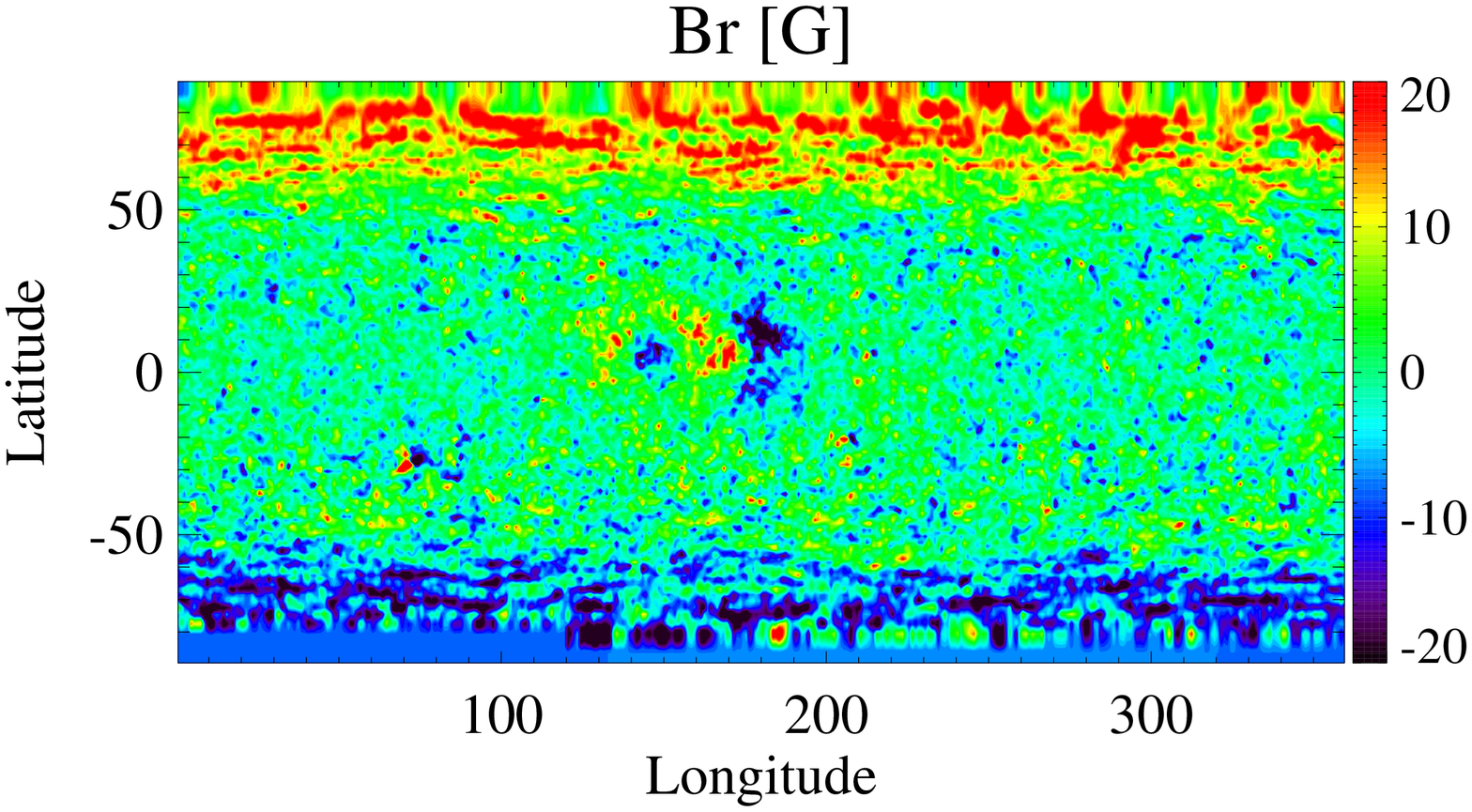}{0.5\textwidth}{\vspace{-1.0cm}(b) CR2219 NSO-HMI-NRT map}}
\vspace{-1.0cm}
\gridline{\fig{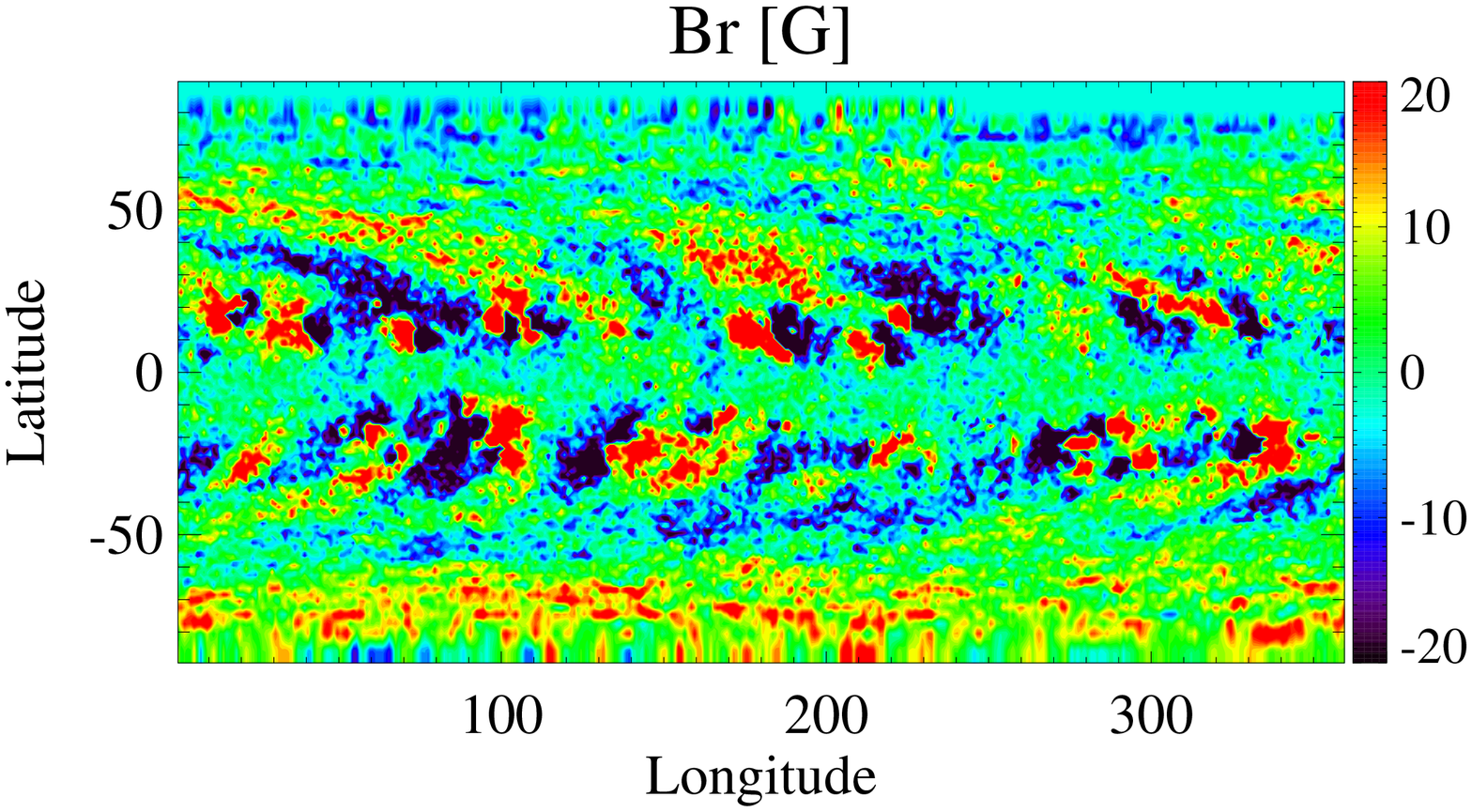}{0.5\textwidth}{\vspace{-1.0cm}(c) CR2123 NSO-HMI-NRT map}
\fig{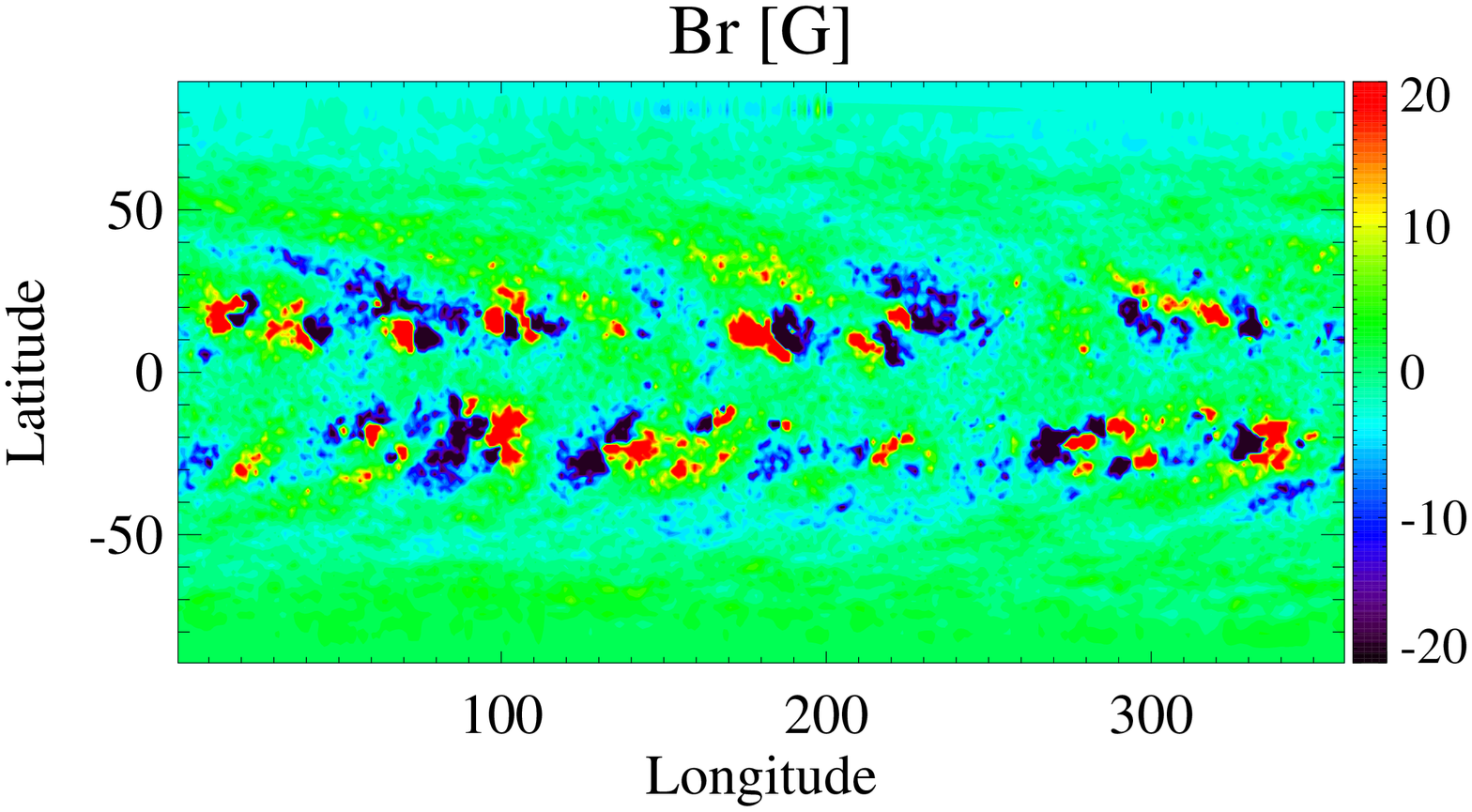}{0.5\textwidth}{\vspace{-1.0cm}(b) CR2123 GONG map}}
\vspace{-0.8cm}
\caption{NSO-HMI-NRT and GONG synoptic maps showing the observed radial photospheric magnetic field. The $B_r$ component from NSO-HMI-NRT maps are shown for Carrington Rotations 2107, 2219 and CR2123 in panels a, b, and c respectively. Panel d shows the $B_r$ field from GONG synoptic map for CR2123. The $B_r$ field range of $\pm$ 20 G is chosen to highlight the features on the map.}\label{fig:maps}
\end{figure*}

AWSoM has been meticulously validated by comparing the simulated results with a variety of observations. Near the Sun, the modeled density and temperature structure of the solar corona is compared to extreme ultraviolet (EUV) observations from \textit{Solar-Terrestrial Relations Observatory} (STEREO, \citet{Howard:2008}), \textit{Solar Dynamics Observatory} (SDO, \citet{Pesnell:2012})/\textit{Atmospheric Imaging Assembly} (AIA, \citet{Lemen:2012}) and \textit{Solar and Heliospheric Observatory}(SOHO)/\textit{Large Angle and Spectrometric Coronagraph} (LASCO, \citet{Brueckner:1995}). In the low corona, AWSoM results have been compared with the tomographic reconstructions of electron density and temperature using EUV and visible-light observations \citep{Lloveras:2017,Lloveras:2020,Lloveras:2022,Vasquez:2022}. In the inner heliosphere, AWSoM successfully reproduces the velocity observations of InterPlanetary Scintillation (IPS) data \citep{Jackson:1998} and the solar wind plasma parameters at 1 au (WIND observations) \citep{Jin:2017a,Sachdeva:2019}). AWSoM has been successful in simulating observed solar wind properties during both solar minimum and maximum conditions \citep{Sachdeva:2019,Sachdeva:2021}.

\section{NSO-HMI-NRT maps}\label{sec:maps}
Here, we discuss the methodology of creating the NSO-HMI-NRT maps used in this paper for prescribing the magnetic field at the AWSoM inner boundary. Synoptic maps are constructed over a full solar rotation by adding new observations of the solar disk as they rotate into the observer's view. Assuming that the Sun rotates as a solid body, the synoptic maps cover the entire solar surface over a whole Carrington Rotation ($\sim$ 27.27 days) \cite[for a general description see, e.g.][Section 7]{Pevtsov:2021}. Here we limit the discussion of charts representing full vector magnetic field, B$_r$ (radial, or up-down), B$_\varphi$ (zonal, East-West), and B$_\theta$ (meridional, North-South).

Various factors contribute to uncertainties in the synoptic charts. Posing the largest challenge are the limited contemporaneous observations of the solar surface, particularly in the polar regions. 
Instrumental noise, conditions of observations and similar factors are also inherited in the maps. 
Smearing of solar features due to differential rotation is also a potential issue, particularly when creating high-spatial resolution synoptic maps. Because the differential component of solar rotation increases with latitude, the smearing effect will mostly be prominent within the polar regions (above 60 degrees).
A detailed description of this problem and a possible solution is given in \cite{2006SoPh..235...17U}.
However, due to the low resolution of the synoptic maps used in this
study, this correction is not required.

As a first step in the creation of a synoptic map, full disk images are remapped from sky-plane coordinates to heliographic coordinates. If remapping is required, the image resolution is reduced to match the resolution of the synoptic map (e.g., 1 by 1 degree in solar latitude and longitude). Next, these sampled or remapped images are added to a synoptic map based on heliographic coordinates of pixels. One method used by SDO/HMI team employs vertical strips of about 2 solar degrees wide centered at the solar central meridian. It produces the so-called diachronic maps. This method is quick and easy since it does not require excessive additional processing. However, it requires sufficiently high cadence in observations as the strips are simply added one after the other, similar to a picket fence. Therefore, any gap in observations results in a gap in the diachronic map. Furthermore, such a map fails to correctly represent any features that emerge or drastically evolve after passing the central meridian. 

The NSO-HMI-NRT synoptic maps used in this paper are generated using a different approach which incorporates the use of the full disk SDO/HMI magnetograms to build a synoptic map \citep{Bertello:2014}. This method can be computationally expensive if the cadence of full-disk magnetograms is too high. However, the maps created with this technique include all the magnetic features regardless of when they appear during a rotation or whether they evolve significantly before and/or after passing the central meridian.
Each NSO-HMI-NRT magnetic field map in this paper incorporates approximately 43 (8+27+8) days of observations. That is, in addition to the 27 days of a Carrington rotation, the synoptic maps cover eight days (each) before and after the rotation. While adding 8 days before/after the start/end of a Carrington rotation is not necessary, it allows for a better equalisation of weight (or number of contributing full disk observations) for each heliographic pixel. Without this contribution, the first (last) 8 days of a Carrington rotation map will see a gradual increase (decrease) in a normalized number of contributing points from just a few percent at the leading (trailing) edge to 100\% in the center of the map. Because of this difference in weights, without additional 8 days, the noise level would be slightly higher for the beginning and end parts of each map. A similar procedure is adopted in creating the synoptic maps of pseudo-radial field using GONG observations. Our past experience with HMI vector observations have shown that the observation on the 48$^{th}$ minute of every hour provides the best coverage and data quality for the one-hour cadence that is used here. The selection of the 48$^{th}$ minute is not critical, and has no impact on the results of our project. Nevertheless, it may yield the synoptic maps of a slightly better quality. 

The SDO/HMI data is acquired from the Joint Science Operations Center (JSOC, \url{http://jsoc.stanford.edu/}). This process begins by using a custom python program utilizing an http ‘get request’ to the JSOC server to query the data that is available for download at that given time.  This request is used to verify and record the availability status of all five Data Record Management System (DRMS) segments per observation that can be used to build a given synoptic map. Within JSOC these segments are identified as field, inclination, azimuth, disambig, and conf$\_$disambig (map of the confidence in each pixels disambiguation) for the Full-Disk Milne-Eddington inversion data series (hmi.B$\_$720s), each covering 720 s of observation. A one-hour cadence or 5 segments per hour are used to generate a synoptic map. If all five DRMS segments are available, they are separated into 5 lots of up to 9 days of data or $9\times24\times5=1080$ segments each. However, full block availability is rare and usually a few gaps of missing data occur every few days.  Even with these gaps, this sums up to about 70 GB of observational data (before processing). This large amount of data requires parallelized workloads of each lot to reduce the computational time needed for the next steps. The data acquisition is followed by pre-processing of the SDO/HMI images of the photospheric magnetic field into a single coordinate system transformed package.

Pre-processing includes the coordinate transformation from the image (sky) plane to 
heliographic (solar latitude-longitude) coordinates, re-imaging the full disk data to larger pixels used for construction of a synoptic map, and applying $\cos^4$ of central meridian distance weighting function. For additional details see \citet{Bertello:2014,Hughes.etal2016}. After the pre-processing, the data are used to assemble a complete synoptic map by averaging the contribution to a corresponding synoptic map pixel from all contributing pre-processed images.

\begin{figure*}[ht!]
\centering
\gridline{ \fig{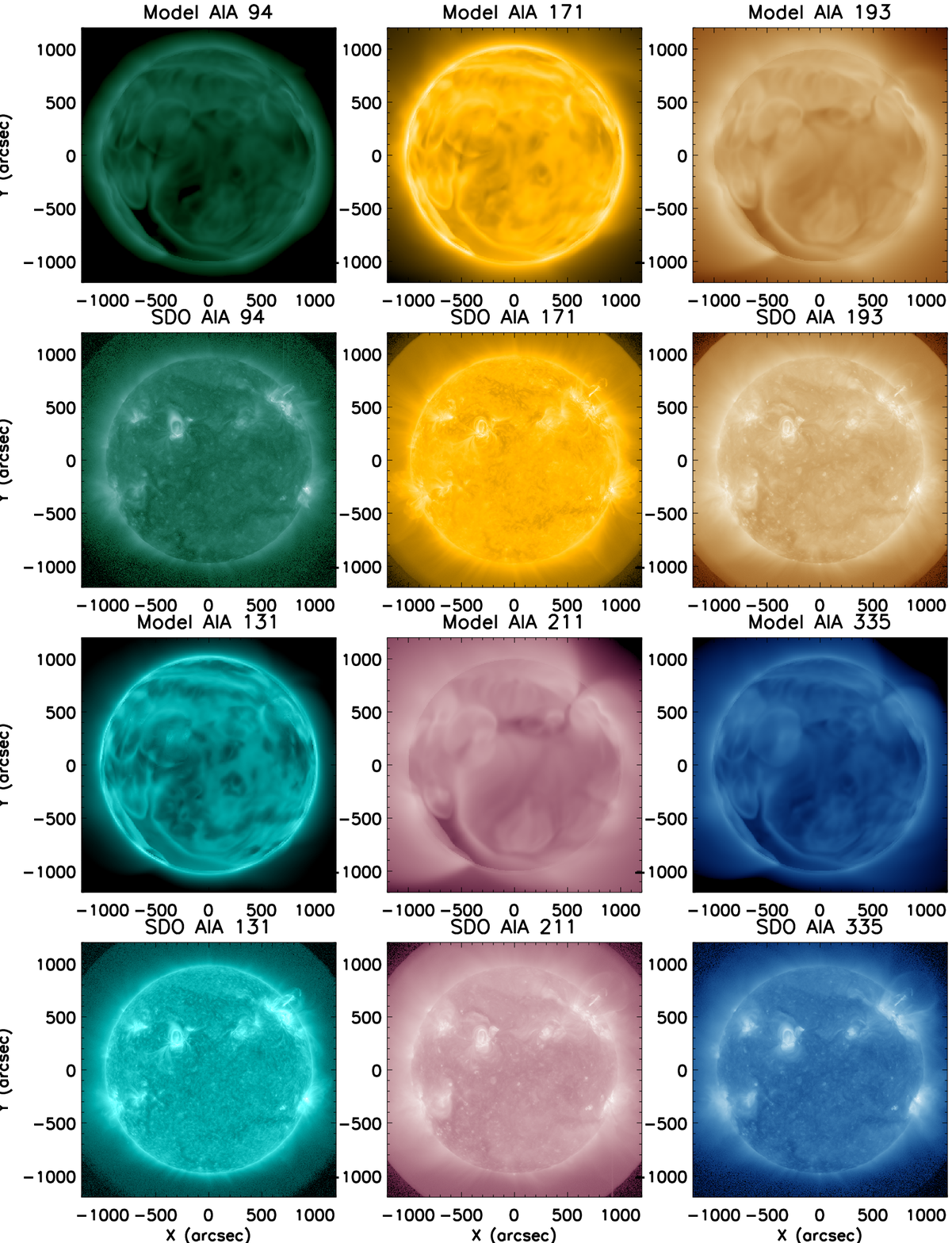}{0.425\textwidth}{(a) CR2107 with NSO-HMI-NRT map}
\fig{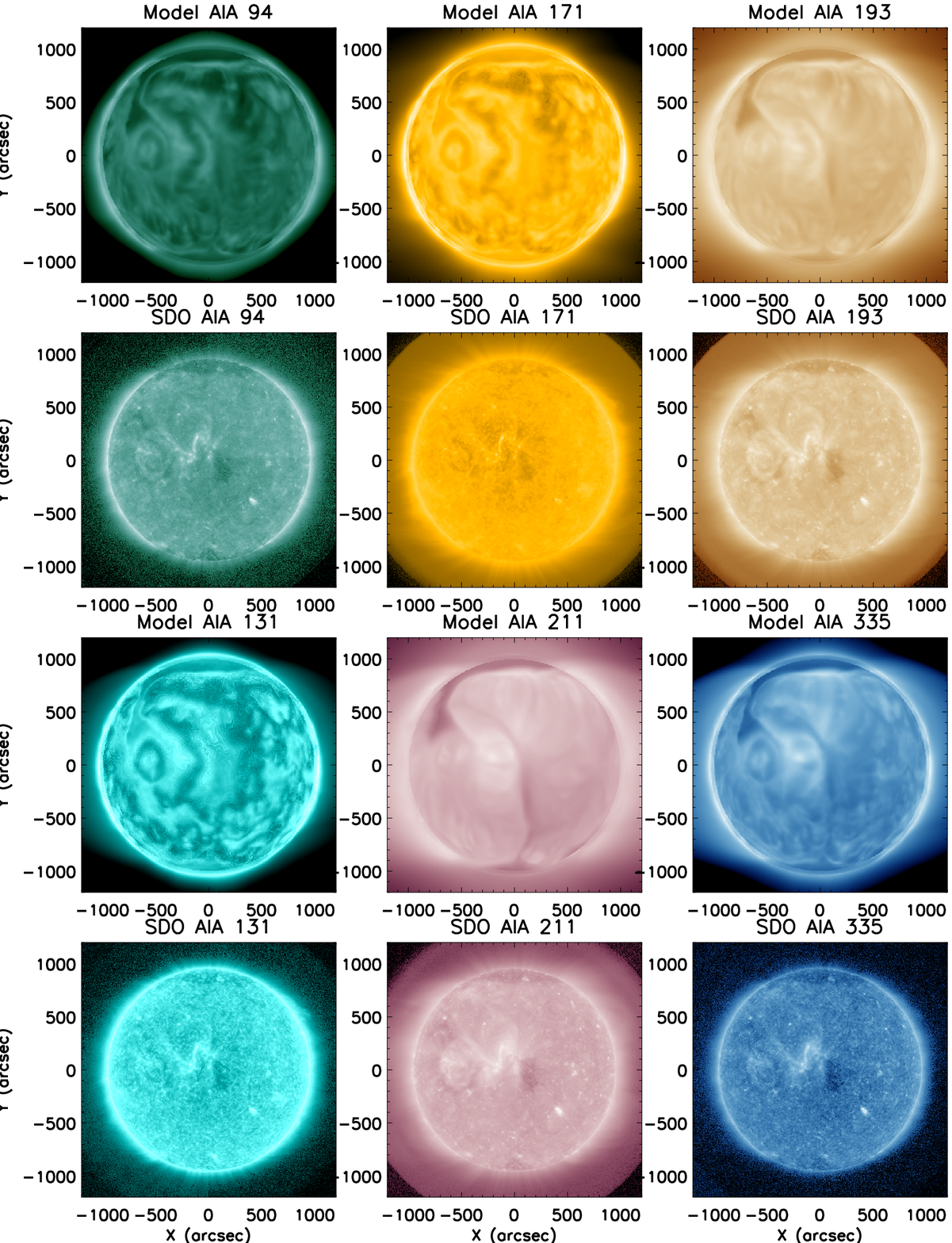}{0.425\textwidth}{(b) CR2219 with NSO-HMI-NRT map}}
\vspace{-0.5cm}
\gridline{\fig{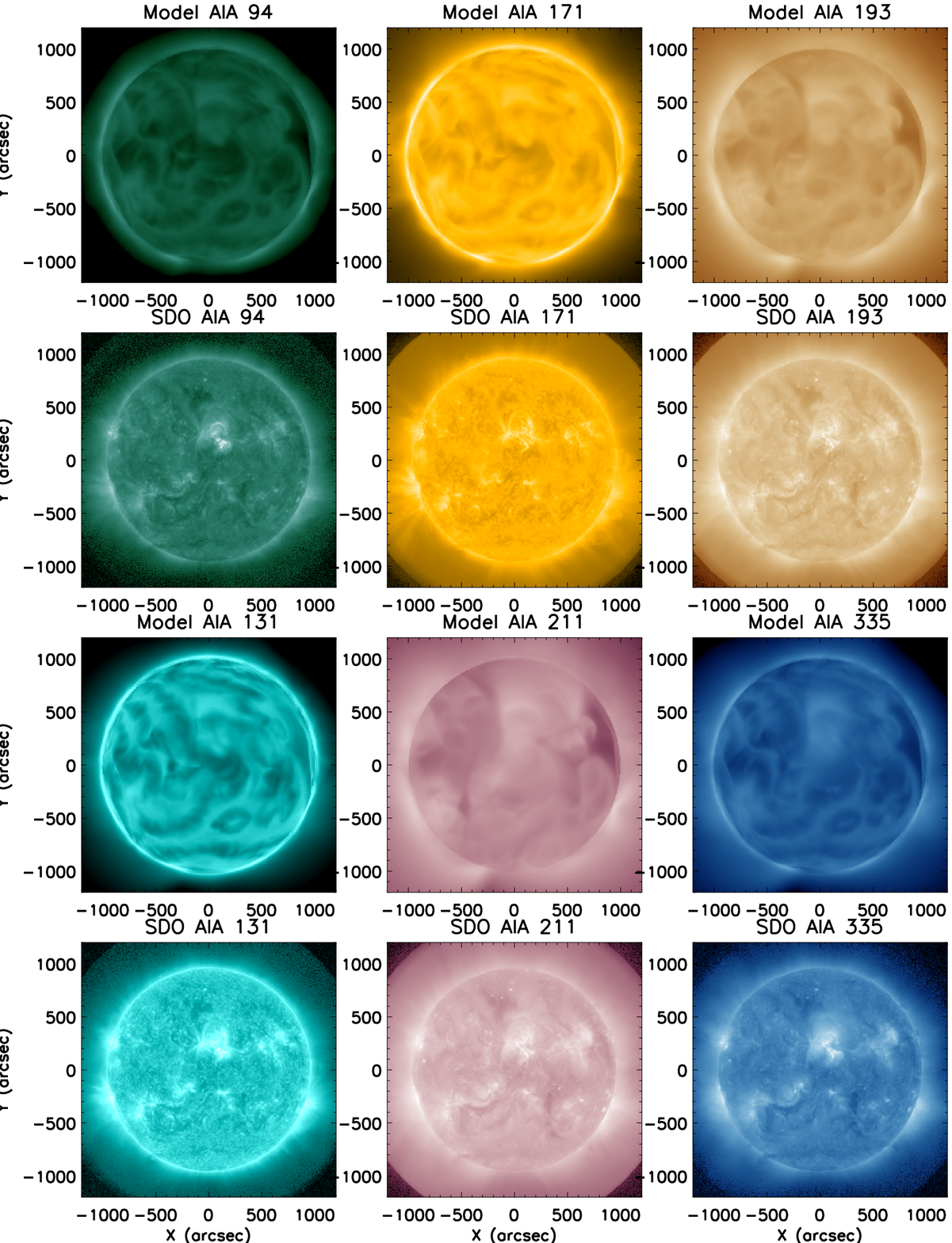}{0.425\textwidth}{(c) CR2123 with NSO-HMI-NRT map}
\fig{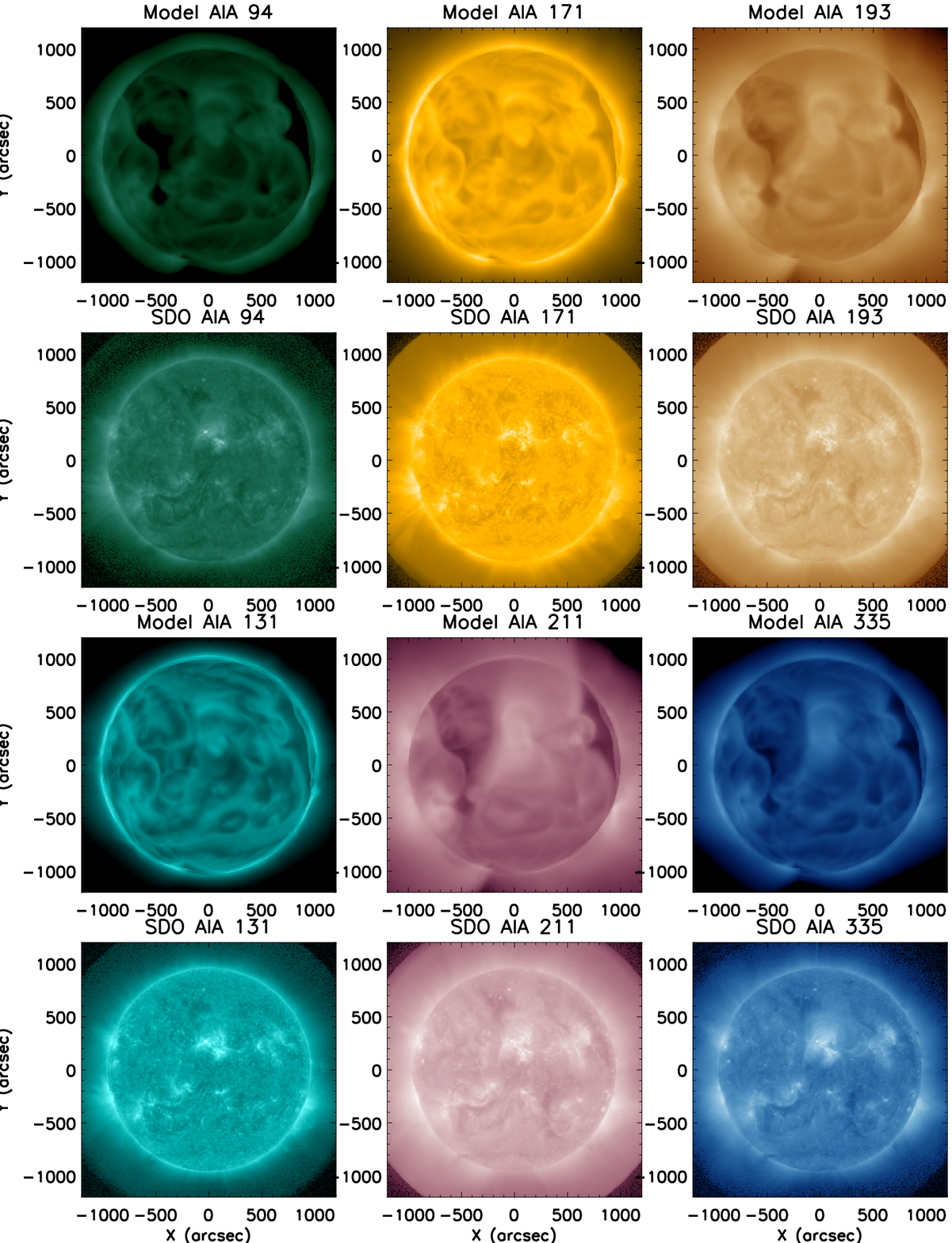}{0.425\textwidth}{(d) CR2123 with GONG map}}
\caption{Comparison of synthetic EUV images with SDO/AIA observations. Panels a, b and c show model-data comparison for CR2107, CR2219 and CR2123 modeled using the NSO-HMI-NRT maps. Panel d shows the same for CR2123 modeled using GONG magnetogram. In each panel, the first and third rows represent the modeled AIA output and the second and fourth rows show the SDO/AIA observations. The comparison is shown in six wavelength channels (94, 171,193, 131, 211 and 335 \AA).} \label{fig:LOS}
\end{figure*}

A well-known problem in constructing a full-surface synoptic map is the limited visibility of the polar fields of the Sun from near the Earth. The tilt angle between the Earth's ecliptic plane and the solar rotation axis varies between about $\pm 7.25^{\circ}$ each year. The poles can therefore only be observed from the ecliptic plane with a large ($> 80^{\circ}$) viewing angle. Moreover, each pole is not observable from near-Earth for more than six months in a year. The unobserved polar fields are therefore required to be modeled. A simple approach, adopted here, is to fill the pixels corresponding to the unobserved polar data using a cubic 
surface fit to the observational data from the neighboring latitudes. 
Numerous studies suggest that the polar fields are approximately radially directed \citep{Svalgaard.etal1978,Petrie:2015}. \citet{Ulrich.Tran2013} argued for a slight $\approx$ 6$^\circ$ poleward inclination of magnetic field in polar areas. \citet{Pevtsov:2021} applied a similar technique, and found about 3$^\circ$ equatorward inclination when using SOLIS/VSM data. \citet{Virtanen.etal2019} also found a small ($\le$10$^\circ$) equatorward inclination at high ($\sim$75$^\circ$) latitudes. Based on the results of these previous studies, we set the unobserved $B_{\theta}$ and $B_{\phi}$ values in the polar regions to zero and fill B$_r$ using cubic surface fit. Finally, this process provides a complete synoptic map.

\section{Numerical Simulation Set-up} \label{sec:setup}
We use the NSO-HMI-NRT magnetic field maps for three Carrington rotations (CRs) to drive the background solar wind simulations. The solar corona (SC) and inner heliosphere (IH) components of the SWMF are used via AWSoM (Section \ref{sec:awsom}). The NSO-HMI-NRT synoptic maps used this in study for Carrington rotations CR2107 (2011-02-16 to 2011-03-16), 
CR2123 (2012-04-28 to 2012-05-25), 
and CR2219 (2019-06-29 to 2019-07-26) are shown in panels a, b and c of Figure \ref{fig:maps}.
Each of these maps show the structure of the photospheric magnetic field obtained from the SDO/HMI images followed by the procedure described in the previous Section \ref{sec:maps}. 
In a previous work, \citet{Jin:2017a} used the \textit{Global Oscillation Network Group} (GONG) synoptic magnetogram to simulate the solar wind conditions for CR2107 using AWSoM and compared the simulation results with OMNI data at 1 au and along the trajectory of STEREO-A. \citet{vanderHolst:2014awsom,Meng:2015} showed an improvement in the results for this rotation when they use the magnetic field obtained from SDO/HMI instrument \citep{Scherrer:2012}.

The 2D photospheric magnetic field from the synoptic maps is used to reconstruct the 3D magnetic field using the Potential Field Source Surface Model (PFSSM). The radial component of the observed magnetic field is used as the boundary condition for the potential solver while the longitudinal and latitudinal components are allowed to relax to a solution. We use the spherical harmonics solution for the PFSSM with the source surface at 2.5 \Rs. At the inner boundary, the initial temperature for both isotropic electron and perpendicular and parallel proton temperature is set to 50,000 K. The proton number density at these temperatures is overestimated to provide a ready source to replenish the plasma, which maybe depleted due to chromospheric evaporation \citep{Lionello:2009,vanderHolst:2014awsom}.
AWSoM has a limited number of free parameters that may be varied to improve the results when compared to observations of the solar wind. The energy density of the outward propagating \alf waves is set using the Poynting Flux ($S_{A}$) of the wave which is proportional to the magnetic field strength at the inner boundary (B$_{\odot}$) \citep{Fisk:2001a,Fisk:1999b,Sokolov:2013}. A recent study by \citet{Huang:2022} using AWSoM showed that the quantity \PF needs to be varied based on the phase of the solar cycle.
During phases of stronger magnetic activity, the amount of energy of the outward propagating \alf wave is reduced by reducing the \PF parameter to avoid deposition of excess energy density into the chromosphere and high density peaks at 1 au \citep{Sachdeva:2021}. For CR2107, CR2123 and CR2219 the optimal value of the \PF parameter in the model is set to 0.35, 0.3 and 1.0 in units of $10^{6}$ Wm$^{-2}$T$^{-1}$, respectively. The \alf wave correlation length (L$_{\perp}$), which is transverse to the magnetic field direction is proportional to B$^{-1/2}$ \citep{Hollweg:1986} and is set to $1.5\times10^{5}$ m\,$\sqrt{T}$.

The SC component uses a 3D spherical grid extending from 1 - 24 \Rs and is coupled with the IH component which uses a Cartesian grid that extends from -250 to 250 \Rs. The SC and IH components are coupled with a buffer grid extending from 18-21 \Rs to transfer the SC solution to the IH domain in the steady-state run. The SC domain is decomposed into 6$\times$8$\times$8 grid blocks while IH has 8$\times$8$\times$8 grid blocks. The computation includes Adaptive Mesh Refinement (AMR) in SC which provides an angular resolution of 1.4$^{\circ}$ below 1.7 \Rs and 2.8$^{\circ}$ in the remaining domain. The number of cells in SC and IH are 4.2 million and 12.2 million respectively. The cell size in IH ranges between 0.48 \Rs and 7.8 \Rs. Using local time stepping, the SC component is run for 80000 iterations and coupled with IH for one step followed by 5000 steps in IH to get the steady state solution. Additional AMR is done below 1.7 \Rs along with the 5th order shock-capturing scheme \citep{Chen:2016} to produce high resolution line of sight synthetic EUV images for comparison with observations.

\begin{figure*}[th!]
\centering
\gridline{ \fig{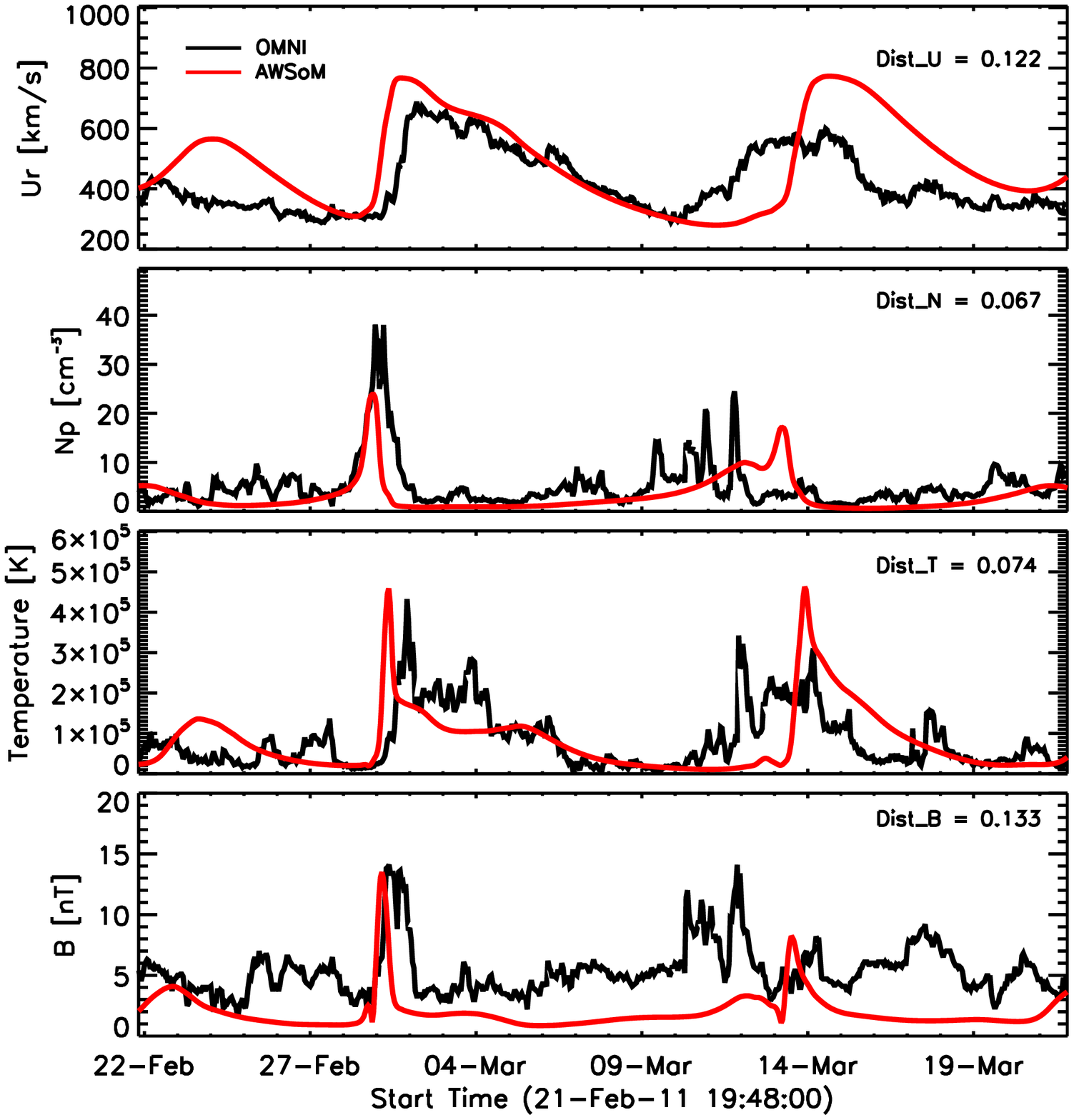}{0.45\textwidth}{(a) CR2107 with NSO-HMI-NRT map}
           \fig{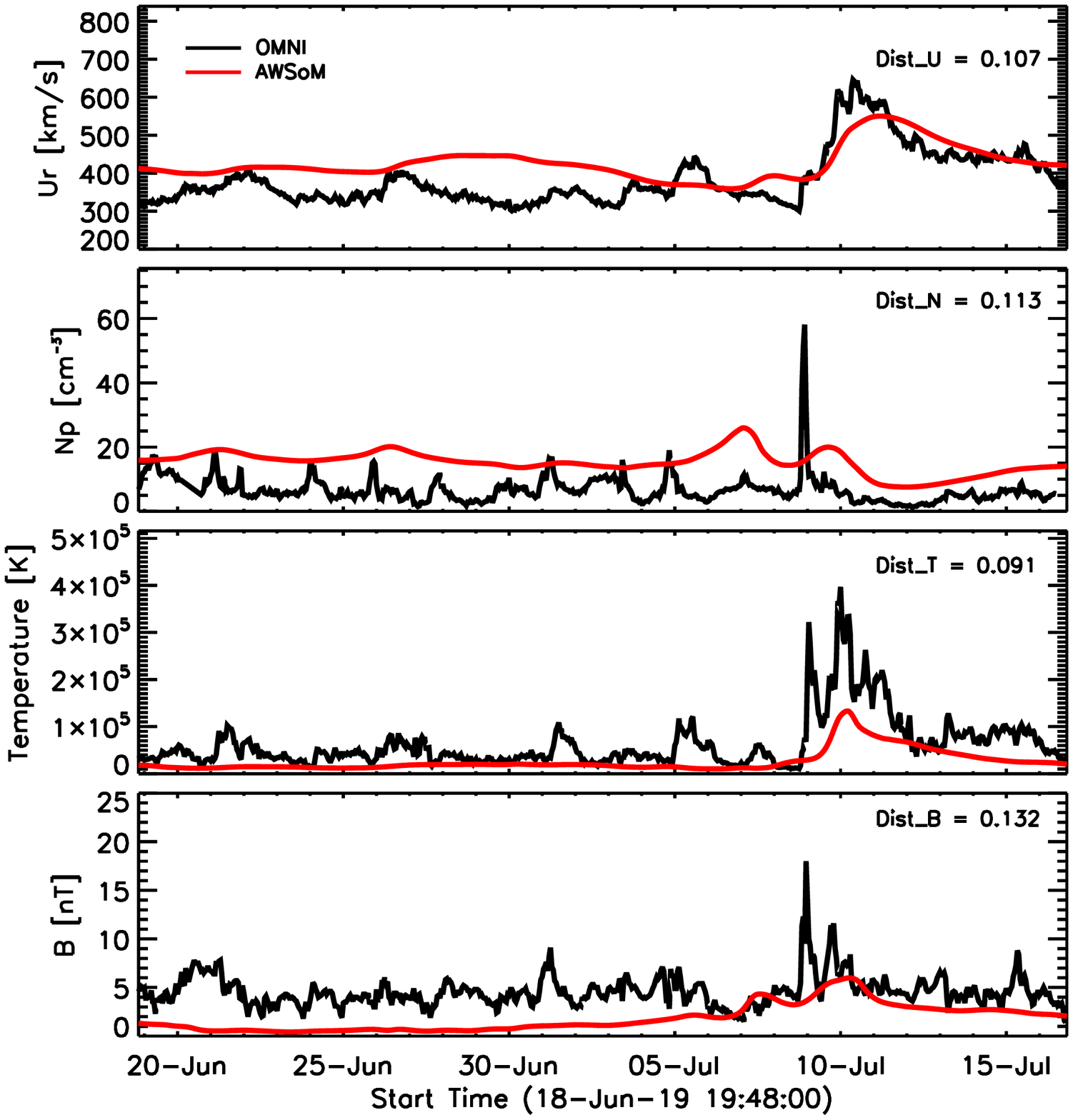}{0.45\textwidth}{(b) CR2219 with NSO-HMI-NRT map}}
\gridline{\fig{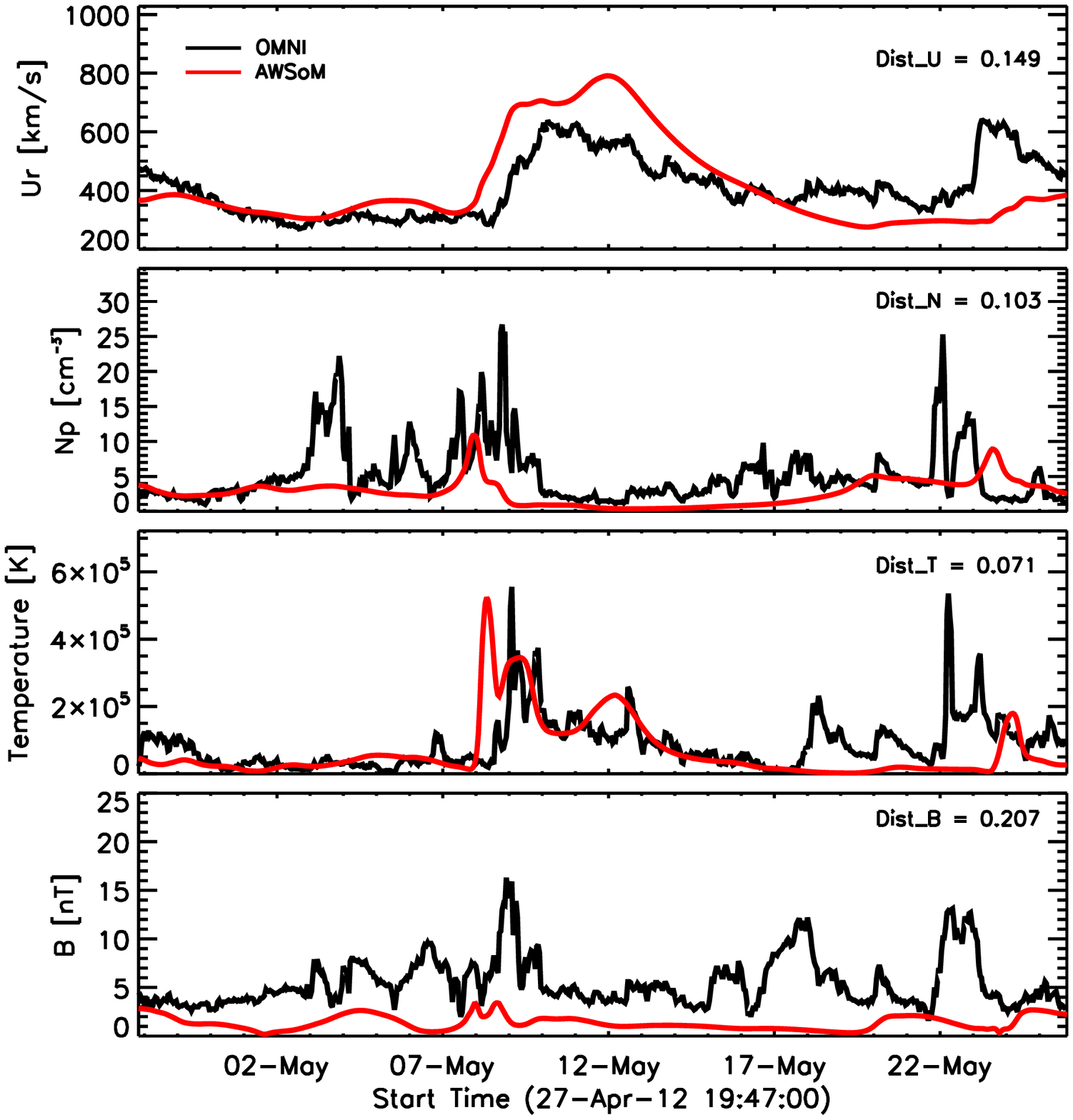}{0.45\textwidth}{(c) CR2123 with NSO-HMI-NRT map}
           \fig{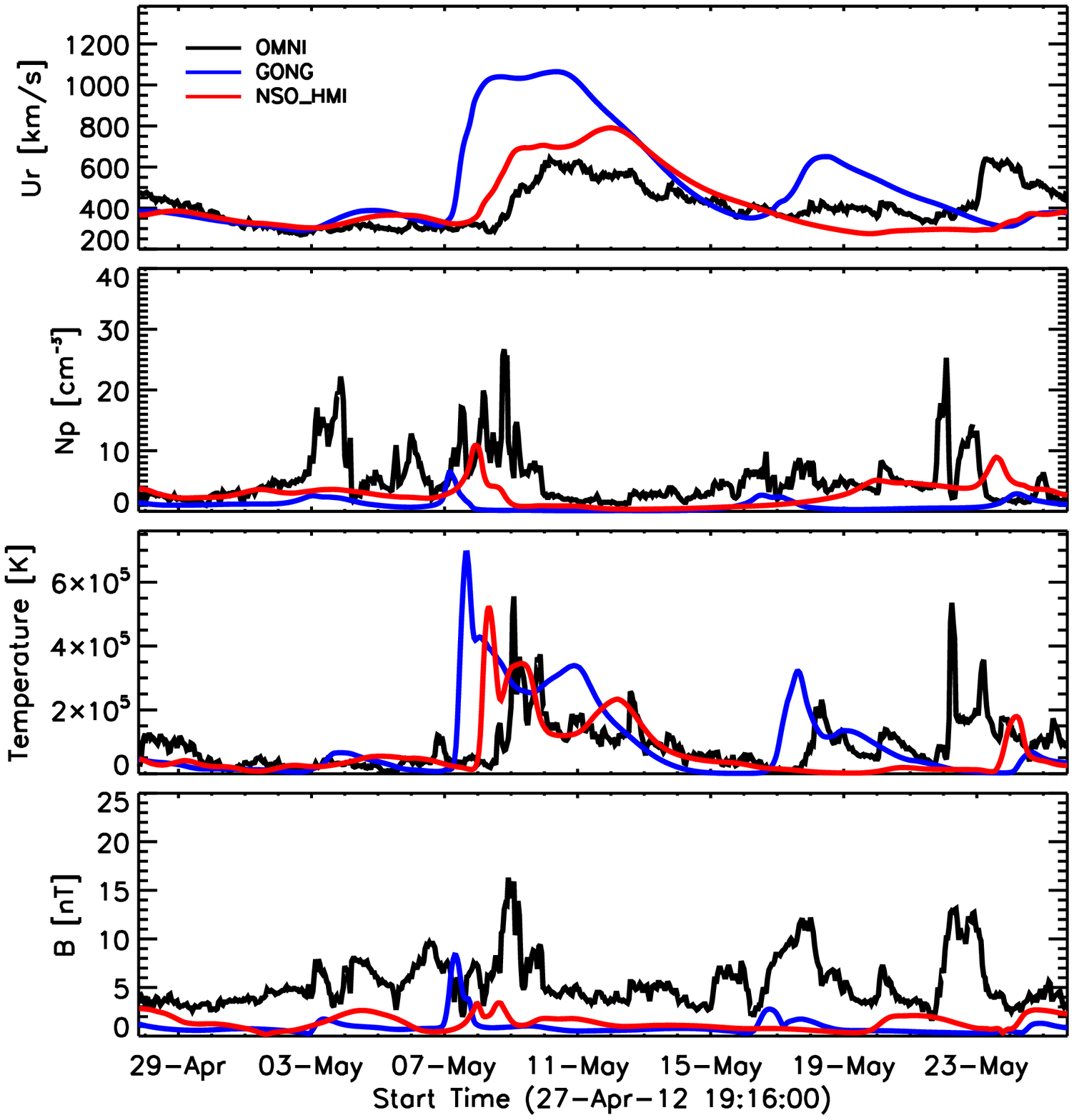}{0.45\textwidth}{(d) CR2123 with GONG and NSO-HMI-NRT maps}}
\caption{Comparison of AWSoM simulated 1 au solar wind plasma parameters with the 1-hr averaged OMNI observations for the three CR's. Model results are in red and data is in black. Panels a, b and c correspond to data-model comparisons for CR2107, CR2219 and CR2123 respectively where the simulations were driven by the NSO-HMI-NRT maps (shown in Figure \ref{fig:maps}). Panel d shows the comparison of simulation results from AWSoM driven by the NSO-HMI-NRT (red) and GONG (blue) maps for CR2123.
}\label{fig:OMNI}
\end{figure*}

\section{Results} \label{sec:results}
We use the magnetic field from the NSO-HMI-NRT maps to obtain steady-state solar wind solutions for three CRs 2107, 2123 and 2219. Figure \ref{fig:maps} shows the NSO-HMI-NRT synoptic maps depicting detailed features of the active regions as well as the polar regions as described in Section \ref{sec:maps}. These maps represent different phases of the solar cycle. CR2107 and CR2123 correspond to higher solar activity with stronger magnetic field and more active regions in comparison to solar minimum conditions found in CR2219. Panels c and d of Figure \ref{fig:maps} show the comparison between the magnetic field maps from NSO-HMI-NRT (left) and GONG (right) for CR2123. More small-scale features are present in the NSO-HMI-NRT map as well as stronger magnetic fields in the active regions. In the GONG map, the polar magnetic fields are weaker and smoother, a distinct difference which will impact the speed of the modeled solar wind.

The simulation domain for AWSoM covers the low corona making it ideal to obtain synthetic extreme ultraviolet (EUV) images that can be compared with corresponding observations. 
Figure \ref{fig:LOS} shows the synthetic line-of-sight images from the AWSoM simulation results compared with corresponding SDO/AIA observations in six wavelength channels (94, 171, 193, 131, 211 and 335 \AA). Here, panels a, b and c show model-data comparisons for CR2107, CR2219 and CR2123 respectively, modeled using the NSO-HMI-NRT maps. The first and third row of images in each panel show the model synthesized AIA images and the second and fourth rows of images show the corresponding SDO/AIA observations. Panel d of Figure \ref{fig:LOS} shows the same model-data comparison for CR2123 modeled using the GONG magnetogram. 
The simulation results compares well with observations in matching the overall brightness and the location of the major active regions for the three rotations. This fact suggests that AWSoM can reproduce the 3D structure of the density and temperature in the low solar corona. The modeled coronal holes appear to be darker in comparison to observations but match in their location and extent. Although the average brightness matches well in all channels, the bright active regions can best be seen accurately in 193, 211 and 335 \AA\, channels. For example, for CR2219 (Panel b), during solar minimum, the model AIA images reproduce the major bright active region which can be seen clearly in these wavelength channels. For CR2123, synthetic AIA images obtained from driving the model using the NSO-HMI-NRT and GONG maps show major differences in the overall brightness of the active regions and the coronal holes. In particular, the modeled coronal hole from the GONG-driven simulation appears to be much darker.  
Additional refinement of the grid with the AWSoM model can further improve model comparisons by producing brighter active regions \citep{Shi:2022}.
\begin{figure*}[ht!]
\centering
\gridline{ \fig{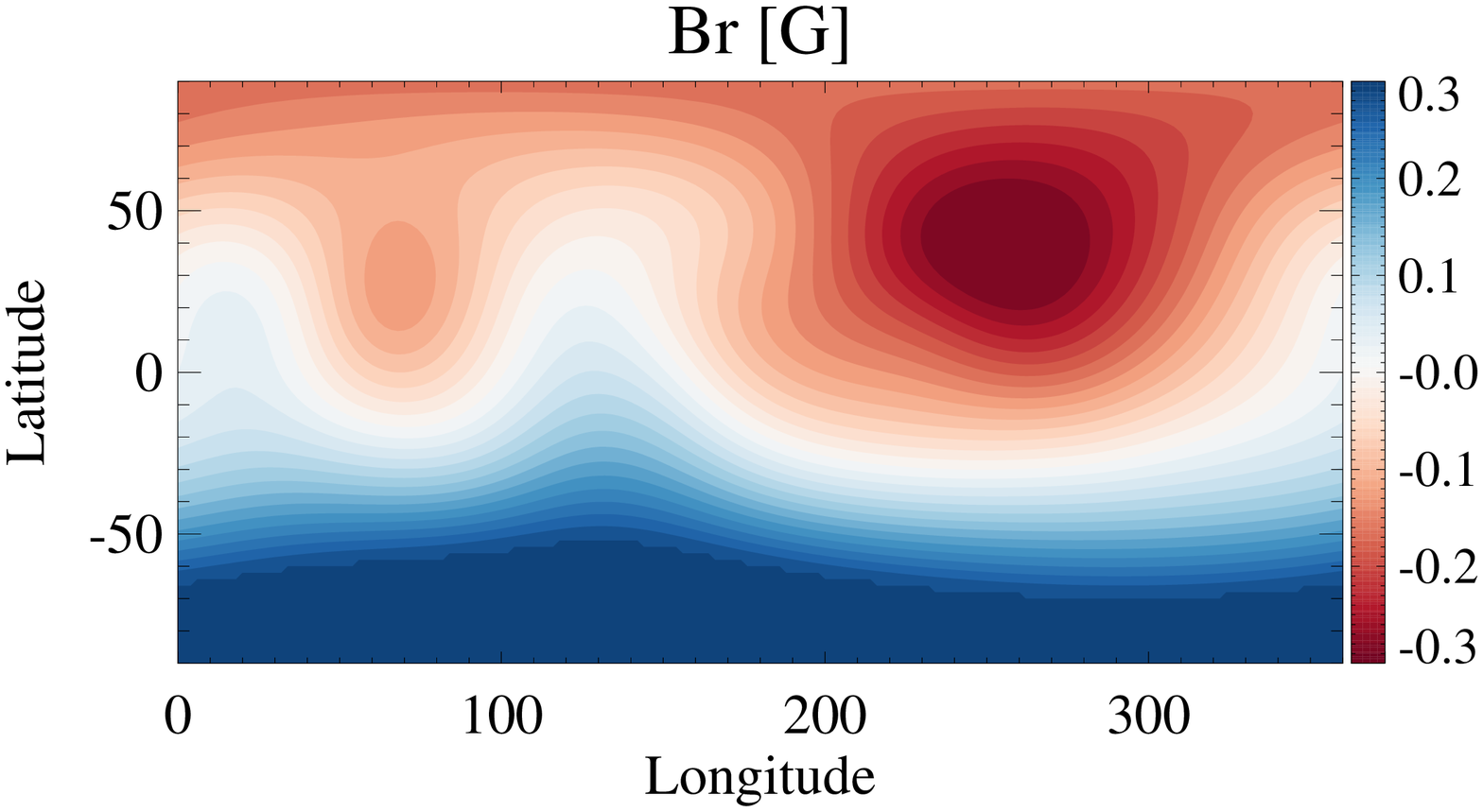}{0.5\textwidth}{\vspace{-1.2cm}(a) CR2107 NSO-HMI-NRT}
           \fig{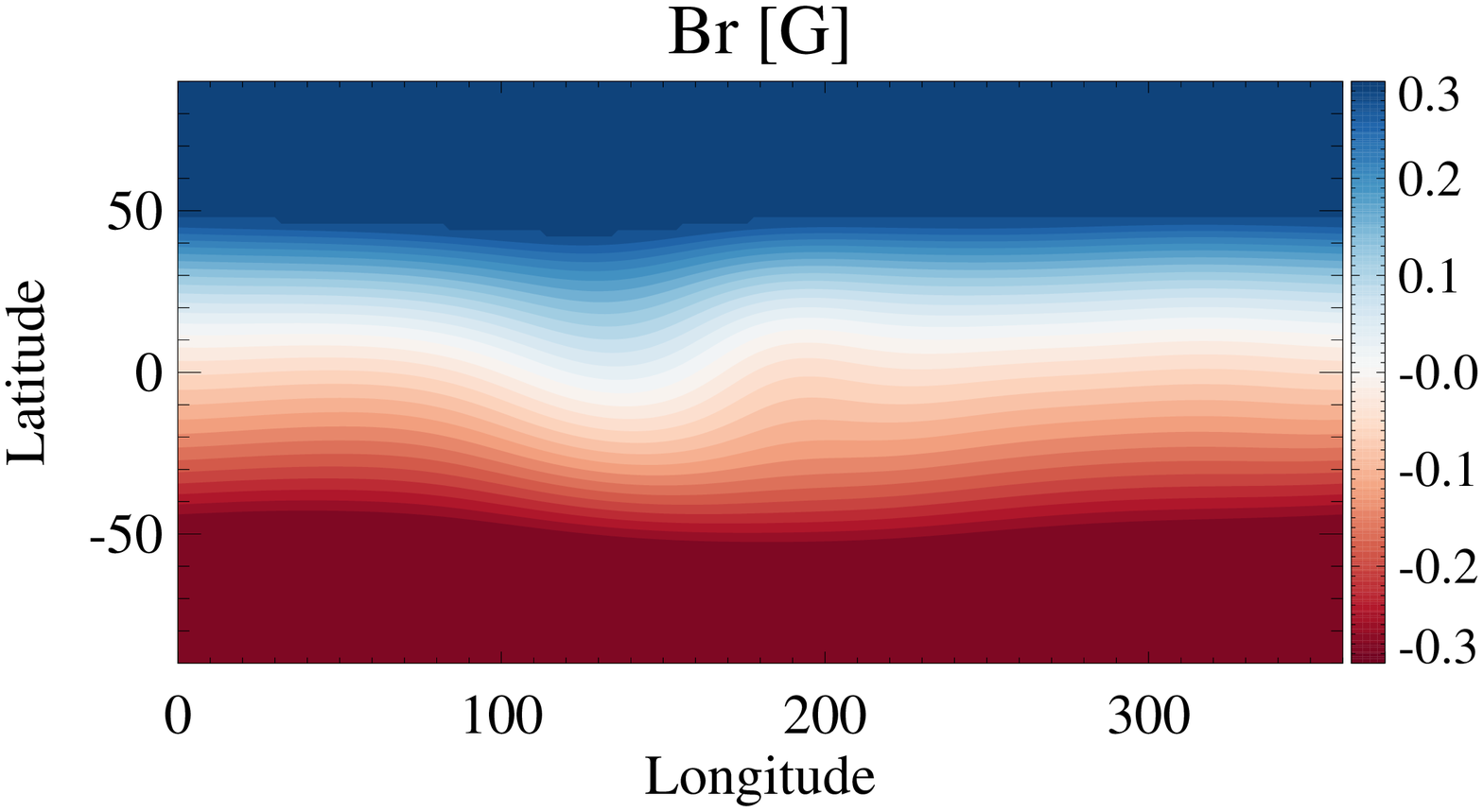}{0.5\textwidth}{\vspace{-1.2cm}(b) CR2219 NSO-HMI-NRT}}
           \vspace{-1.8cm}
\gridline{\fig{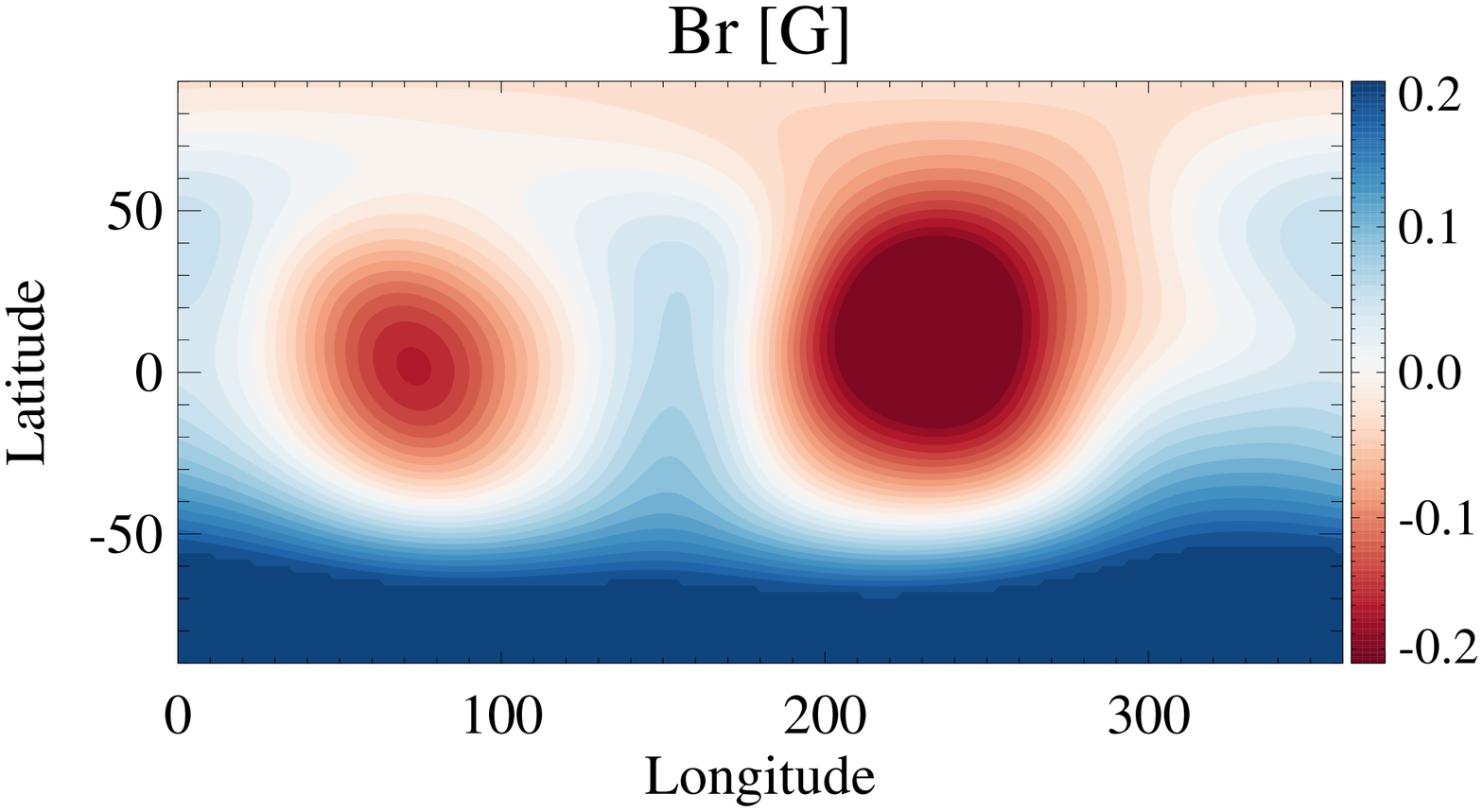}{0.5\textwidth}{\vspace{-1.2cm}(c) CR2123 NSO-HMI-NRT}
           \fig{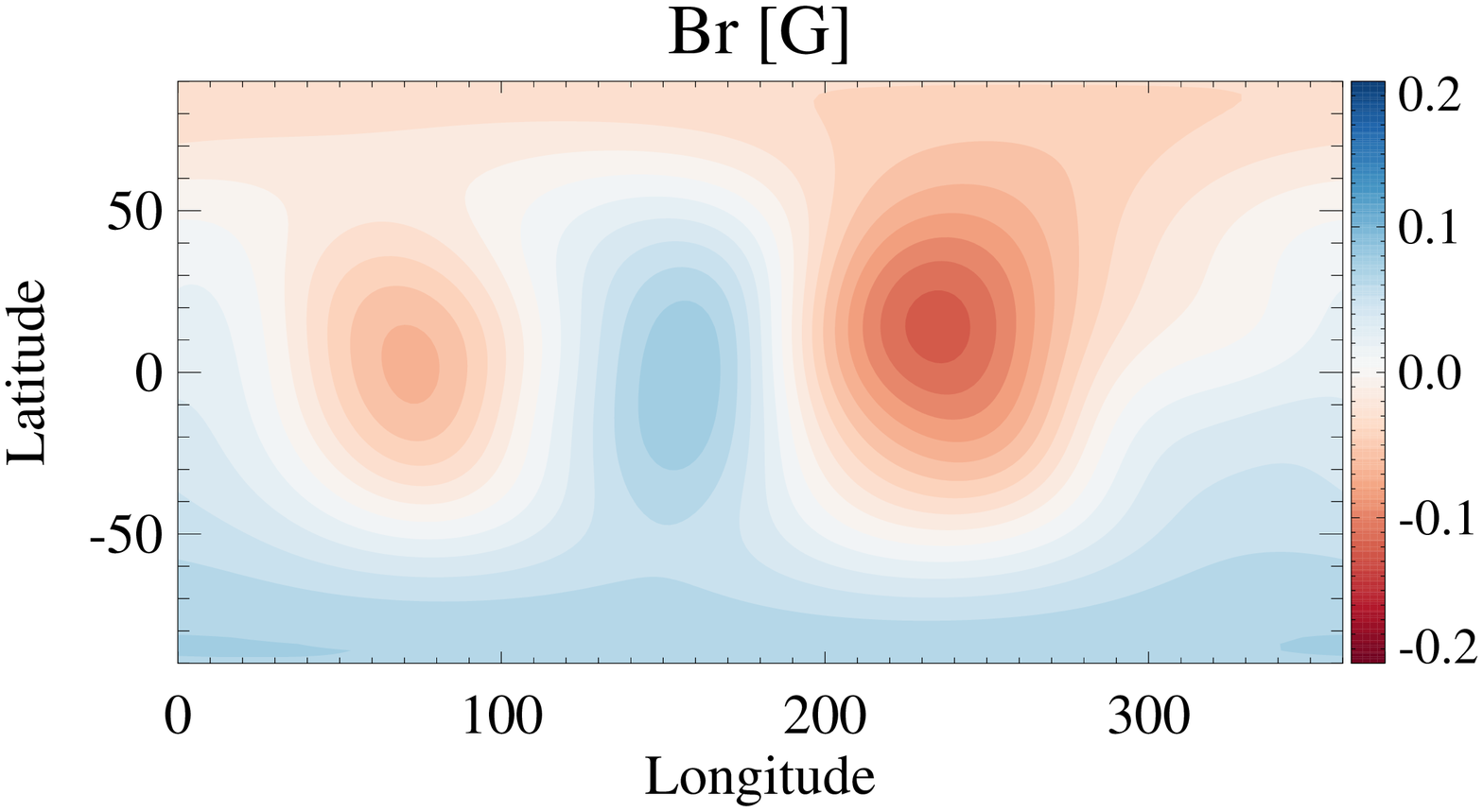}{0.5\textwidth}{\vspace{-1.2cm}(d) CR2123 GONG }}
 \vspace{-0.8cm}    
\caption{Radial magnetic field at the source surface radius (2.5 \Rs). Panels a, b and c show the $B_{r}$ magnetic field for CR2107, CR2219 and CR2123 respectively at 2.5 \Rs calculated from the PFSSM using the NSO-HMI-NRT maps. Panel d shows the same for CR2123 using the GONG synoptic map. The source surface in the PFSSM is set to 2.5 \Rs for all the maps.}\label{fig:PFSS}
\end{figure*}

To compare the simulated solar wind with {\it in situ} observations of plasma parameters at L1 we extract (from the 3D result) the model solution along the trajectory of the Earth. Figure \ref{fig:OMNI} shows the AWSoM output along the Earth's trajectory in red color and the OMNI data in black for all three rotations. We see that overall, the model when driven by NSO-HMI-NRT maps successfully reproduces the observed solar wind plasma. In particular, we see that for both CR2107 and CR2123 the solar wind solution matches quite well with the observations for all quantities. For CR2107, the model predicts the co-rotating interaction region (CIR) on March 1, 2011. The solution matches the significant jump in the radial speed $(U_r)$, proton density $(N_p)$, ion temperature and the absolute magnetic field $(B)$. For both rotations that represent the near solar maximum phase (CR2107 and CR2123), the features in the solar wind plasma parameters are well matched by the model solution. In both case, however, we find that the magnetic field is under-predicted and the peak model speed is overestimated by about 11 $\%$ and 23 $\%$ for CR2107 and CR2123 respectively. As a result of the higher speed, the CIR in both cases arrive slightly earlier in the model as compared to the observations. CR2219 is a period of reduced activity for which the model overestimates the solar wind speed and density. However, the CIR speed in the model matches well with the observations. To further quantify the model-data comparison, we calculate a distance measure $Dist$ listed in each plot, which informs us of how well the model matches observations. Described in detail in \cite{Sachdeva:2021}, the quantity Dist is a measure of the distance between two curves independent of the coordinates. Smaller values indicate a better fit. 

Panel d of Figure \ref{fig:OMNI} shows the OMNI observations in black and the model results for CR2123 driven by NSO-HMI-NRT and GONG maps (Panels c and d of Figure \ref{fig:maps}) are shown in red and blue color, respectively. The modeled solutions differ significantly at 1 au, which is a direct result of the different initial magnetic field conditions from the two maps. All other model parameters are kept the same for both simulations. This demonstrates that the observational magnetic field input driving the solar corona models significantly impacts the solar wind properties.

Figure \ref{fig:PFSS} represents the radial magnetic field (B$_{r}$) at the source surface radius of 2.5 \Rs obtained from the PFSSM using spherical harmonics with order 180 for each of the rotations. 
In panels a, b and c, the field B$_{r}$ is obtained from the NSO-HMI-NRT maps for CRs 2107, 2219 and 2123. For comparison, panel d shows the $B_{r}$ field obtained from the GONG map for CR2123. Both maps for CR2123 (Panels c and d) are shown on the same scale to highlight the differences between them. The field obtained from the NSO-HMI-NRT map is much more pronounced in the coronal holes as well as the polar regions in comparison to the GONG map.

To quantify this effect, we also calculate the total unsigned open magnetic flux for all the maps at $2.5$ \Rs. This quantity is an integral of the absolute value of the radial magnetic field, $|B_{r}|$, over the source surface. The total unsigned open magnetic flux at 2.5 \Rs is found to be 10.3, 15.9, and 6.9 [Gauss \Rs$^2$] for the NSO-HMI-NRT maps for CR2107, CR2219 and CR2123, respectively. For the GONG map for CR2123, the total unsigned magnetic flux obtained is 3.1 [Gauss \Rs$^2$] at 2.5 \Rs. The scaling law by \cite{Pevtsov:2003} relates the total unsigned flux to the energy deposition in the solar corona, therefore, a stronger total unsigned open flux leads to more energy, which accelerates and powers the solar wind. In relation to AWSoM, the Poynting flux outgoing into to the solar wind is directly proportional to the unsigned open magnetic flux and the constant ratio of the Poynting flux to the magnetic flux is one of the input parameters of the model \PF. Therefore, 
the stronger open flux from the NSO-HMI-NRT map provides more energy to the corona, which increases chromospheric evaporation, increasing the density of the solar wind while reducing its speed.  The result is a better comparisons with observations at 1 au compared to the model results made with the GONG map for CR2123.

\section{Summary and Discussion} \label{sec:conclude}
In this work, we show the impact of the magnetic field conditions obtained from the NSO approach of creating near-real-time maps using the HMI magnetic field observations (NSO-HMI-NRT maps) on the modeled solar wind. The methodology used for these maps includes using full-disk HMI magnetogram but with a weighted pixel contribution and the unobserved polar regions are filled using a polynomial fit to neighbouring observations. 

We use the 3D MHD model AWSoM to simulate the Sun to Earth background solar wind for three Carrington rotations (2107, 2219 and 2123). AWSoM is driven by the magnetic field from the NSO-HMI-NRT maps to demonstrate their performance during varying periods of solar activity. We compare the AWSoM simulated solar wind solutions with observations in the low corona and find that for all three CR's modeled using the corresponding NSO-HMI-NRT maps as input, the large-scale properties of the solar corona including the extent and location of coronal holes as well as regions of enhanced activity (active regions) compare well with SDO/AIA observations. 

Further away from the Sun, we compare the observed solar wind properties at 1 au with the model results. The 1 au observations are reproduced reasonably well by the NSO-HMI-NRT map driven AWSoM model for all three rotations. We find that the while magnetic field is underestimated, the solar wind speed, density and CIR properties are reproduced in the simulations.
For one of the rotations (CR2123), we obtain the solar wind solution using the GONG synoptic map as the initial condition for the photospheric magnetic field and compare the results with the NSO-HMI-NRT map driven solar wind conditions. We find that the solutions with the NSO-HMI-NRT map perform well both in the low corona and at 1 au when compared to the GONG map results. Finally, we show the radial magnetic field at 2.5 \Rs from the PFSSM for each of the maps and compare the total unsigned open magnetic flux at the source surface. This quantity for the NSO-HMI-NRT map is larger by a factor of $\approx \,2$ in comparison to the GONG map for the same rotation.

It is well-known that numerical models of the solar corona are sensitive to the observed magnetic field inputs obtained from a variety of synoptic magnetograms available in the community. Here, we highlight the performance of NSO produced, HMI observation based, near-real-time maps (NSO-HMI-NRT) maps with our 3D extended MHD model (AWSoM) and show that the NSO-HMI-NRT maps are a valuable data product allowing for coronal/solar wind simulations of equal or better quality than those obtained by standard synoptic maps such as GONG maps.

\section*{Acknowledgments}
This work was primarily funded by the NASA 80NSSC17K0686 grant. N. Sachdeva, W. Manchester and Z. Huang were also supported by NASA LWS grant 80NSSC22K0892 and NSF grants PHY-2027555 and 1663800. W. Manchester was also supported by NASA LWS grant NNX16AL12G. I. Sokolov was also supported by NSF -2149771.

Original full disk magnetograms that were used to create synoptic maps are courtesy of NASA/SDO and the HMI science team. The National Solar Observatory (NSO) is operated by the Association of Universities for Research in Astronomy (AURA), Inc., under a cooperative agreement with the National Science Foundation. 
We acknowledge the high-performance computing support from: a) Cheyenne (doi:10.5065/D6RX99HX) provided by NCAR's Computational and Information Systems Laboratory, sponsored by the NSF, b) Frontera (doi:10.1145/3311790.3396656) sponsored by the NSF and c) the NASA supercomputing system Pleiades.

\end{document}